\documentclass{article}
\pdfoutput=1
\usepackage{arxiv}
\usepackage[utf8]{inputenc} % allow utf-8 input
\usepackage[T1]{fontenc}    % use 8-bit T1 fonts
\usepackage{hyperref}       % hyperlinks
\usepackage{url}            % simple URL typesetting
\usepackage{booktabs}       % professional-quality tables
\usepackage{amsfonts}       % blackboard math symbols
\usepackage{nicefrac}       % compact symbols for 1/2, etc.
\usepackage{microtype}      % microtypography
\usepackage{graphicx}
\usepackage{longtable}
\usepackage{adjustbox}
\usepackage{multirow}
\usepackage{float}
\usepackage{threeparttablex}

\usepackage[listings,skins,breakable]{tcolorbox}
\definecolor{verde_pastel}{RGB}{36, 143, 36}
\definecolor{rojo_oscuro}{RGB}{128, 0, 0}
\usetikzlibrary{calc}

\title{Medical imaging data structure extended to multiple modalities and anatomical regions}

\hypersetup{
pdftitle={Medical imaging data structure extended to multiple modalities and anatomical regions},
pdfsubject={q-bio.NC, q-bio.QM},
pdfauthor={Saborit-torres J.M.},
pdfkeywords={standardisation, extension, Brain Imaging Data Structure, Medical Imaging Data Structure, databases, anatomical regions, modality},
}

\begin{document}
\date{}
\maketitle

\author[a,$\star \star$]{~\href{https://orcid.org/0000-0002-2445-6075}{Jose Manuel Saborit-Torres}\hspace{1mm}\includegraphics[scale=0.06]{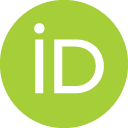};~}
\author[a]{~\href{https://orcid.org/0000-0003-3332-9710}{Jhon Jairo Saenz-Gamboa}\hspace{1mm}\includegraphics[scale=0.06]{orcid.png};~}
\author[a]{~Joaquim Àngel Montell Serrano;~}
\author[a,d]{~\href{https://orcid.org/0000-0003-2520-1178}{Jose María Salinas}\hspace{1mm}\includegraphics[scale=0.06]{orcid.png};~}
\author[i]{~\href{https://orcid.org/0000-0002-4174-3762}{Jon Ander Gómez Adrián}\hspace{1mm}\includegraphics[scale=0.06]{orcid.png};~}
\author[a]{~Ioan Stefan Makay;~}
\author[a]{~Marisa Caparrós Redondo;~}
\author[a,g]{~\href{https://orcid.org/0000-0001-8354-5636}{Francisco García-García}\hspace{1mm}\includegraphics[scale=0.06]{orcid.png};~}
\author[m]{~Julio Domenech Fernández;~}
\author[i]{~Jose Vicente Manjón;~}
\author[h,l]{~\href{https://orcid.org/0000-0002-6228-2678}{Gonzalo Rojas Costa}\hspace{1mm}\includegraphics[scale=0.06]{orcid.png};~}
\author[j]{~\href{https://orcid.org/0000-0002-9445-5529}{Antonio Pertusa}\hspace{1mm}\includegraphics[scale=0.06]{orcid.png};~}
\author[e]{~\href{https://orcid.org/0000-0001-6527-3059}{Aurelia Bustos}\hspace{1mm}\includegraphics[scale=0.06]{orcid.png};~}
\author[c,f]{~German González;~}
\author[d]{~Joaquin Galant;~}
\author[a,b,k,$\star$]{~\href{https://orcid.org/0000-0003-4505-8399
}{María de la Iglesia-Vayá}\hspace{1mm}\includegraphics[scale=0.06]{orcid.png}.}
\begin{center}
    $^{\star}$ Correspondence: \href{mailto:delaiglesia\_mar@gva.com}{delaiglesia\_mar@gva.es}

    $^{\star \star}$ Correspondence: \href{mailto:saborit_jostor@externos.gva.es}{saborit\_jostor@externo.gva.es}

\end{center}
\begin{flushleft}
	[a]{Unidad Mixta de Imagen Biomédica FISABIO-CIPF. Fundación para el Fomento de la Investigación Sanitario y Biomédica de la Comunidad Valenciana - Valencia, Spain.}
	
	[b]{ ~General Directorate of Research, Innovation, Technology and Quality. Subdirectorate General of Information Systems for Health  - València, Spain.}
	
	[c]{\raggedright ~Universidad de Alicante - Alicante, Spain.}
	
	[d]{\raggedright ~Hospital San Juan de Alicante - Alicante, Spain.}
	
	[e]{\raggedright ~Medbravo.}
	
	[f]{\raggedright ~Sierra Research SL.}
	
	[g]{\raggedright ~Bioinformatics \& Biostatistics Unit, Principe Felipe Research Center - Valencia, Spain.}
	
	[h]{\raggedright ~Department of Psychology, Stanford University - California, United States}
	
	[h]{\raggedright ~Laboratory for Advanced Medical Image Processing, Department of Radiology, Clínica las Condes - Santiago, Chile}
	
	[i]{\raggedright ~Department of Computer Systems and Computation, Universitat Politècnica de València - València, Spain}
	
	[j]{\raggedright ~University Institute for Computing Research, University of Alicante - Alicante, Spain}
	
	[k]{\raggedright ~CIBERSAM, ISC III. Av. Blasco Ibáñez 15, 46010 - València, Spain}
	
	[l]{\raggedright ~Health Innovation Center, Clínica las Condes - Santiago, Chile}
	
	[m]{\raggedright ~Hospital Arnau de Vilanova - Valencia, Spain}\vspace{5mm}

\end{flushleft}

	\begin{center}
	Saborit-Torres and Saenz-Gamboa contributed equally to this work.
\end{center}\vspace{5mm}

\begin{abstract}
	Brain Imaging Data Structure (BIDS) allows the user to organise brain imaging data into a clear and easy standard directory structure. BIDS is widely supported by the scientific community and is considered a powerful standard for management. The original BIDS is limited to images or data related to the brain. Medical Imaging Data Structure (MIDS) was therefore conceived with the objective of extending this methodology to other anatomical regions and other types of imaging systems in these areas.
\end{abstract}

% keywords can be removed
\newpage

\section{Introduction}
Methods which yield reliable and reproducible results must be used when acquiring scientific knowledge. High test-retest reliability of the applied methods is the foundation of research, irrespective of the scientific discipline. It is in the prime interest of every scientict to obtain reproducible results. While such reproducibility was considered of utmost importance in the positron emitting tomography (PET) field \cite{adams2010systematic}, the quantitative assessment of reproducibility has largely been neglected in the fMRI community, or as Bennett and Miller described it: “Reliability is not a typical topic of conversation” between functional magnetic resonance imaging (fMRI) investigators \cite{bennett2010reliable}. This situation changed significantly in 2016 following the establishment of the Committee on Best Practices in Data Analysis and Sharing \cite{cobidas2016web}  by the leading neuroimaging society - the Organisation for Human Brain Mapping (OHBM).

The basis of the Valencia Medical Imaging Bank  \cite{bimcv2017web} is the clinical environment data curation  proposal, by which imaging data can be collected correctly and efficiently. Finding a way to organise this information is crucial. A proper organization and curation of the images is essential to train deep learning methods \cite{lecun2015deep} that can perform object detection and segmentation using reliable medical data. Metadata can also be included in multimodal classifiers to complement imaging data in order to improve the accuracy of the detection.

Images and medical information can be stored in different ways, although there is no standard that indicates how this information should be organised and shared. The Health Ministry’s Centre of Excellence and Innovation for Image Technology recommends using a simple system so that any researcher can understand the data distribution \cite{ceib2014web}.

The proposed structure is called MIDS (Medical Imaging Data Structure) and aims to be a new standard that contains all types of medical information and images in simple hierarchical folders. It was conceived as an extension of the standard Brain Imaging Data Structure \cite{bids2016web,gorgolewski2016brain}, which stores brain images. MIDS takes this system further and is not confined to brain images only. The idea is to create the same structure for images of different body parts by magnetic resonance, computed tomography, ecography, etc, following a single process, regardless of the type and shape of the image.

\section{Methodology}
\label{sec:methodology}

\subsection{Brain Imaging Data Structure}

Many studies focus on obtaining a medical imaging dataset for their own purposes, so that the management and control of the associated images and metadata can be roughly an effort. During projects, more data are generated and it may be necessary to relocate it inside a dataset. Each study has its own manner of organising the data, which makes it more difficult to understand, while a curated and well structured dataset can improve the search user experience and the quality of automatic classifiers.

\begin{figure}[]
	\centering
	\includegraphics[width=1\linewidth]{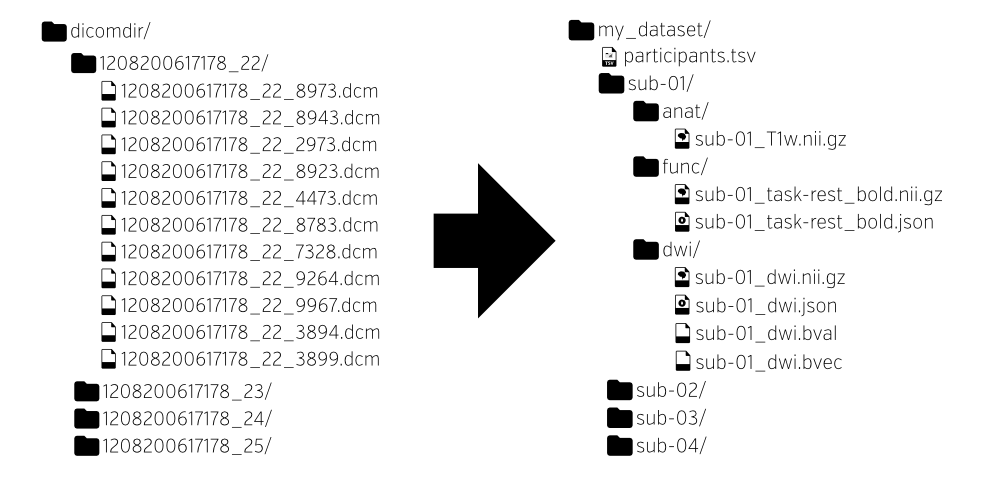}
	\caption{DICOM to BIDS conversion with Dcm2Bids software}
	\label{fig:fig1}
\end{figure}

There are a couple of studies which propose a standard to store this type of data, including BIDS, which aims a standard form of storing magnetic resonance imaging data and metadata in a clear and simple hierarchical folder structure. It is supported by several programs and libraries dedicated to the study of medical images (e.g. c-pacs, freesurfer, XNAT, BIDS Validator, among others) and is widely used by research groups. Figure \ref{fig:fig1} gives an example of the BIDS structure; the left directory is a folder with DICOM (Digital Imaging and Communication On Medicine) images \cite{mildenberger2002introduction} and the right directory is its corresponding BIDS structure.

\newpage
\subsection{Medical Imaging Data Structure}

As BIDS only supports brain MRIs, if a project needs, for example, a lumbar MRI, BIDS would not support the images. However, by expanding its structure, other imaging techniques can be integrated in it, which is how MIDS was created. BIDS is thus a potential standard to store MRIs and there is in practice little difference between BIDS and MIDS. Furthermore, in epidemiological studies based on Population Image, MIDS can incorporate any type of \cite{dicomMod2019c73111} medical image (e.g. Computed Radiography, Computed Tomography, Ultrasound, Mammography, etc). MIDS can thus be seen as an extension of BIDS with a similar structural format \cite{bidsEspecification2019web}.

\subsubsection{General template for other anatomical regions}

\begin{figure}[]
	\centering
	\includegraphics[width=0.7\linewidth]{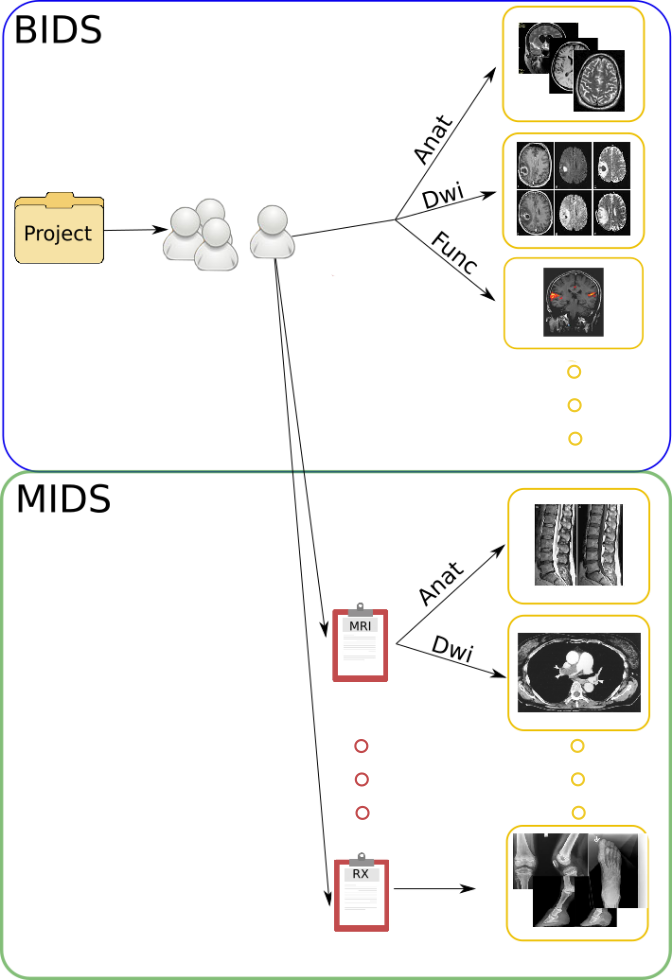}
	\caption{BIDS/MIDS Scheme. MIDS structure adds an extra directory level with the technique used.}
	\label{fig:fig2}
\end{figure}

MIDS adds a new level to the BIDS directory hierarchy which describes the types of medical images used for a particular session. As can be seen in Figure \ref{fig:fig2}, the structure is compatible with BIDS.

The added level is to define the type of medical imaging and can be classified by the energy used for their acquisition, together with the functional or tomographic adjectives for their generation.  The classification is shown in Table \ref{tab:byenergy}.

\begin{table}[H]
\centering
\caption{Classification by the energy used in the acquisition along with its functional or tomographic adjectives.\cite{salinas2013cloud}}
\label{tab:byenergy}
\resizebox{\textwidth}{!}{%
\begin{tabular}{lllll}
\hline
Modality of medical image           & Technique                                         & Energy         & Functional & Tomography \\ \hline
\multirow{3}{*}{General Radiology}  & radiography                                       & X-rays         & No         & No         \\
                                    & Radioscopy                                        & X-rays         & No         & No         \\
                                    & Computed Tomography, CT                      & X-rays         & No         & Yes        \\
 &   & &  &       \\ 
\multirow{2}{*}{Nuclear Medicine}   & Single Photon Emission Computed Tomography, SPECT & $\gamma-$rays         & Yes        & Yes        \\
                                    & Positron Emission Tomography, PET                 & $\gamma-$rays         & Yes        & Yes        \\ 
 &   & &  &       \\ 
Ultrasound                          & Ultrasound                                        & ultrasound     & No         & Yes        \\ 
 &   & &  &       \\ 
\multirow{2}{*}{Magnetic Resonance} & Magnetic resonance imaging, MRI                   & radiofrequency & No         & Yes        \\
                                    & radiofrequency                                    & radiofrequency & Yes        & Yes        \\ 
 &   & &  &       \\ 
Endoscopy                           & Endoscopy                                         & light          & No         & No         \\ 
 &   & &  &       \\ 
Microscopy                          & Microscopy                                        & light          & No         & Yes        \\ \hline
\end{tabular}%
}
\end{table}

\subsubsection{File tags in MIDS}

This template includes a new level to describe other types of medical image than MRI. The researcher decides whether or not to use particular filename keys, depending on the type of medical image. For example, the \textit{contrast enhancement} (ce-$<$\textit{label}$>$) filename key can be used for CT but will not be necessary for Bone Densitometry (X-ray).

The image types supported by MIDS, their relationship  DICOM and their corresponding label can be seen in Table 2. The keys can be described as follows:

\begin{itemize}
    \item acq-$<$\textit{label}$>$: denotes the set of acquisition parameters used (defined in BIDS - optional).
    \item rec-$<$\textit{label}$>$: denotes the reconstruction used; “norm” means normalised images (defined in BIDS - optional).
    \item run-$<$\textit{index}$>$: denotes repetition of identical acquisition with identical scanning parameters (defined in BIDS - optional).
    \item bp-$<$\textit{BodyPartExamined\_label}$>$: denotes the Defined Terms for Body Part Examined in DICOM tag (0018,0015) \cite{dicomMod2019L} (defined in MIDS - optional).
    \item vp-$<$\textit{viewPosition\_label}$>$: which describes the section, view, planes, direction or projection in the acquisition (defined in MIDS - optional). 
    
    Possible labels:
    \begin{itemize}
        \item Planes: \textbf{sag} $=$ Sagittal plane, \textbf{cor} $=$ Coronal plane, \textbf{ax} $=$ Axial plane.
        \item projections: \textbf{ap} $=$ Anterior/Posterior, \textbf{pa} $=$ Posterior/Anterior, \textbf{ll} $=$ Left Lateral, \textbf{rl} $=$ Right Lateral, \textbf{rld} $=$ Right Lateral Decubitus, \textbf{lld} $=$ Left Lateral Decubitus, \textbf{rlo} $=$ Right Lateral Oblique, \textbf{llo} $=$ Left Lateral Oblique.
    \end{itemize}
    \item $[$mod-$]$$<$\textit{modality\_dicom / techniques\_label / modality\_label}$>$: Type of equipment that acquired the original data used to create the images in this Series. DICOM tag (0008,0060) \cite{dicomMod2019c73111}. The key \textit{mod} is used when the type is not the end of the file name. (defined in BIDS and extended in MIDS - required).
    
    \item Extension file:  In BIDS, the extension of image files are NIfTi format with an optional compression of the data (.nii$[$.gz$]$). NIfTi is the optimal format  for 3D neuroimaging data, yet MIDS has types of 2D image data and is not optimal for NIfTi format. This means that 2D images must be saved in Portable Network Graphics (PNG) format \cite{boutell1997png}, which is the best choice because it is a graphical format based on a lossless compression algorithm for non-patent bitmaps (defined in BIDS and extension in MIDS - required)
    
\end{itemize}
\subsubsection{MRI  modality label}

This label refers to anatomical data acquired for a participant. When working with brain imaging, the modalities correspond to the intrinsic values of the resonance machines or sequences currently supported by BIDS (Table \ref{tab:modality_label}).

\begin{table}[!h]
\centering
\caption{The modalities currently admitted in BIDS.}
\label{tab:modality_label}
\resizebox{\textwidth}{!}{%
\begin{tabular}{@{}lll@{}}
\toprule
Name               & modality\_label & Description                                                                                                                                                                                                \\ \midrule
T1 weighted        & T1w             &                                                                                                                                                                                                            \\
T2 weighted        & T2w             &                                                                                                                                                                                                            \\
T1 Rho map         & T1rho           & Quantitative T1rho brain imaging (\cite{johnsonThedens2015, johnson2015}) \\
T1 map             & T1map           & quantitative T1 map                                                                                                                                                                                        \\
T2 map             & T2map           & quantitative T2 map                                                                                                                                                                                        \\
T2*                & T2star          & High resolution T2* image                                                                                                                                                                                  \\
FLAIR              & FLAIR           &                                                                                                                                                                                                            \\
FLASH              & FLASH           &                                                                                                                                                                                                            \\
Proton density     & PD              &                                                                                                                                                                                                            \\
Proton density map & PDmap           &                                                                                                                                                                                                            \\
Combined PD/T2     & PDT2            &                                                                                                                                                                                                            \\
Inplane T1         & inplaneT1       & T1-weighted anatomical image matched to functional acquisition                                                                                                                                             \\
Inplane T2         & inplaneT2       & T2-weighted anatomical image matched to functional acquisition                                                                                                                                             \\
Angiography        & angio           &                                                                                                                                                                                                            \\ \bottomrule
\end{tabular}%
}
\end{table}
\subsubsection{sequences and MIDS modalities}

MR data can be acquired with different extrinsic parameter values, such as echo time, flip angle, inversion time during the same scan.  When we have a population clinical image set, we can find many types of sequences that depend on the manufacturers of the machine used, called extrinsic values of the machine. These sequences can be fundamental since they can produce changes in the resulting image to highlight some tissues in the resonances or improve the quality or speed with which the image is obtained. This is an important feature to be considered when training artificial intelligence methods.

In some cases, these sequences are widely used depending on the area to be observed and the problem to be addressed. We will consider these widely used sequences as modalities. Table \ref{tab:sequences1} gives some examples.

% Please add the following required packages to your document preamble:
% \usepackage{graphicx}
\begin{table}[H]
\centering
\caption{Widely used sequences considered as MIDS modalities.}
\label{tab:sequences1}
\begin{adjustbox}{max width=\textwidth}
\resizebox{1.4\textwidth}{!}{%

\begin{tabular}{lcl}
\toprule
\textbf{Name}                         & \textbf{Techniques\_label} & \textbf{Description}                                                                                                                                                                                                                                                                                                                   \\
\midrule
Short-TI Inversion Recovery           & stir                       & \begin{tabular}[c]{@{}l@{}}It is typically used to nullify the signal \\ from fat. Fat suppression is generally \\ uniform and relatively independent of \\ magnetic field inhomogeneities.\end{tabular}                                                                                                                                 \\
                                      & \multicolumn{1}{l}{}       &                                                                                                                                                                                                                                                                                                                                        \\
HASTE/SS-FSE                          & haste                      & \begin{tabular}[c]{@{}l@{}}SSFSE or HASTE sequence is one of the \\ ultrafast sequences that enables us to\\ acquire the whole MRI data (k-space) in a \\ single radiofrequency (RF) excitation or single shot. Used for  \\ patients for  myelography and routine liver \\ protocol.\end{tabular}                                                            \\
                                      & \multicolumn{1}{l}{}       &                                                                                                                                                                                                                                                                                                                                        \\
Magnetisation transfer                & T1mt, T2mt                 & \begin{tabular}[c]{@{}l@{}}Magnetisation transfer imaging (MTI) applies\\ RF energy exclusively to \\ the bound pool using specially designed MT \\ pulse(s).\\ The relative difference in signal between \\ two adjacent tissues (A and B) is known as \\ the magnetisation transfer contrast (MTC).\end{tabular} \\
                                      & \multicolumn{1}{l}{}       &                                                                                                                                                                                                                                                                                                                                        \\
Dynamic Contrast Enhanced             & DCE                        & \begin{tabular}[c]{@{}l@{}}T1-weighted images following contrast agent \\ injection.\end{tabular}                                                                                                                                                                                                                                       \\
                                      & \multicolumn{1}{l}{}       &                                                                                                                                                                                                                                                                                                                                        \\
Chemical Exchange Saturation Transfer & CEST                       & \begin{tabular}[c]{@{}l@{}}Sensitive images to specific molecules following \\ selective saturation of mobile protons in chemical \\ exchange with water.\end{tabular}                                                                                                                                                                 \\
                                      & \multicolumn{1}{l}{}       &                                                                                                                                                                                                                                                                                                                                        \\
Perfusion                             & pwi                        & \begin{tabular}[c]{@{}l@{}}Comparatively, this establishes the amount of blood \\ received by a certain area of the brain.\end{tabular}   \\ \bottomrule                                                                                                                                                                                                
\end{tabular}%

}
\end{adjustbox}
\end{table}

Since a dataset must have all the information organised and this information must be relevant for further research, the type of sequence must be indicated. We recommend using the \textit{acq} tag to indicate twhich sequence has been used to acquire the image. It should also be indicated in the tabular file of the scan information stipulated in BIDS. See examples in Table \ref{tab:sequences2}.
% Please add the following required packages to your document preamble:
% \usepackage{booktabs}
% \usepackage{longtable}
% Note: It may be necessary to compile the document several times to get a multi-page table to line up properly
\begin{longtable}[c]{@{}lcl@{}}
\caption{Some sequences commonly obtained in magnetic resonance imaging.}
\label{tab:sequences2}\\
\toprule
\textbf{Name}                                                                                                             & \textbf{Sequence\_label} & \textbf{Description}                                                                                                                                                                                                                                                                                                                                                                                                                               \\* \midrule
\endhead
\bottomrule
\endfoot
\endlastfoot
Spin Echo (SE)                                                                                                            & se                       & Standard sequence for MRI                                                                                                                                                                                                                                                                                                                                                                                                                          \\
                                                                                                                          & \multicolumn{1}{l}{}     &                                                                                                                                                                                                                                                                                                                                                                                                                                                    \\
Fast Spin-Echo (FSE)                                                                                                      & fse                      & \begin{tabular}[c]{@{}l@{}} Fast spin echo (FSE) imaging is modification of conventional SE \\using multiple 180º-pulse with different phase-encoding gradients.\end{tabular}                                                                                                                                                           \\
                                                                                                                          & \multicolumn{1}{l}{}     &                                                                                                                                                                                                                                                                                                                                                                                                                                                    \\
\begin{tabular}[c]{@{}l@{}}Gradient  Echo  Pulse  \\ Sequences (GRE)\end{tabular}                                         & gre                      & \begin{tabular}[c]{@{}l@{}}These Maintain better signal-to-noise ratio (SNR) than FLASH at\\ short TR times and are therefore preferred for breath-holding\\ techniques, for example.\\ Used in Thoracic / Cervical trauma  protocol, Cervical  myelo\\ protocols.\end{tabular}                                                                                                                                                                \\
                                                                                                                          & \multicolumn{1}{l}{}     &                                                                                                                                                                                                                                                                                                                                                                                                                                                    \\
\begin{tabular}[c]{@{}l@{}}Spoiled  Gradient  \\ Echo  (SPGR)\end{tabular}                                                & spgr                     & \begin{tabular}[c]{@{}l@{}}Groups  of  sequences  used  to  create  mainly  T1 weighted \\ imaging.  “Spoil”    refers  to  the  sequence  design  feature \\ used  to    crush    or    spoil    any  remaining  magnetisation \\ at  the  end  of  each  TR  cycle.\\ Appear in  Imaging  Protocols for   Internal  Acoustic  Canal,\\  usually   tinnitus,  hearing  loss, acoustic neuroma, or \\ protocol for parotid glands.\end{tabular} \\
                                                                                                                          & \multicolumn{1}{l}{}     &                                                                                                                                                                                                                                                                                                                                                                                                                                                    \\
\begin{tabular}[c]{@{}l@{}}Balanced  Steady-State \\ Free  Precession  (SSFP)\end{tabular}                                & ssfp                     & \begin{tabular}[c]{@{}l@{}}SSFP sequences that can be called GRE, SPGR, Turbo FLASH,\\  and FFE. Some  manufacturers  name  the  fully  balanced  \\ SSFP  sequences as FIESTA, True FISP, or b-FFE.\\ It can be used successfully for Non Contrast MR Angiography \\ (MRA), Cardiac MRI, Magnetic Resonance \\ Cholangiopancreatography (MRCP), Myelography  and \\ other applications.\end{tabular}                                                 \\
                                                                                                                          & \multicolumn{1}{l}{}     &                                                                                                                                                                                                                                                                                                                                                                                                                                                    \\
Phase  Contrast  (PC)                                                                                                     & pc                       & \begin{tabular}[c]{@{}l@{}}Used in Phase contrast MR angiography (PC MRA) to visualise\\ brain arteries and veins by eliminating  signal  from  stationary \\ tissues  in  the  background  using  the  phase information.\end{tabular}                                                                                                                                                                                                            \\
                                                                                                                          & \multicolumn{1}{l}{}     &                                                                                                                                                                                                                                                                                                                                                                                                                                                    \\
Time-of-Flight  (TOF)                                                                                                     & tof                      & \begin{tabular}[c]{@{}l@{}}Time-of-flight  sequences  are  commonly  used  non contrast \\ MRA sequences. The 2D and 3D TOF techniques are used to\\ differentiate the stationary tissue and moving protons inside\\ the blood vessels. It can  be  used  to  visualise  veins,  arteries,\\  or  both (protocols for haemorrhage patients).\end{tabular}                                                                                               \\
                                                                                                                          & \multicolumn{1}{l}{}     &                                                                                                                                                                                                                                                                                                                                                                                                                                                    \\
                                                                                                                          & \multicolumn{1}{l}{}     &                                                                                                                                                                                                                                                                                                                                                                                                                                                    \\
\begin{tabular}[c]{@{}l@{}}Echo  Planar  Imaging \\ (EPI)\end{tabular}                                                    & epi                      & \begin{tabular}[c]{@{}l@{}}The  base  sequence  for  diffusion  weighted  imaging (DWI) \\ sequence in routine pancreas protocol.\end{tabular}                                                                                                                                                                                                                                                                                                      \\
                                                                                                                          & \multicolumn{1}{l}{}     &                                                                                                                                                                                                                                                                                                                                                                                                                                                    \\
Inversion Recovery (IR)                                                                                                   & ir                       & \begin{tabular}[c]{@{}l@{}}Used to suppress signal of a specific tissue or to enhance the\\ contrast in certain applications. With STIR-like sequences, \\ traumatic tissue damage and metastases can be visualised in \\ higher contrast.\end{tabular}                                                                                                                                                                                               \\
                                                                                                                          & \multicolumn{1}{l}{}     &                                                                                                                                                                                                                                                                                                                                                                                                                                                    \\
PROPELLER                                                                                                                 & propeller                & \begin{tabular}[c]{@{}l@{}}Reduces sensitivity to patient's involuntary and physiological\\ movements (respiration, flow, peristalsis) and magnetic\\ susceptibility artifacts.\end{tabular}                                                                                                                                                                                                                                                         \\
\begin{tabular}[c]{@{}l@{}}Driven Equilibrium \\ Single-Pulse Observation \\ of T1\end{tabular}                           & despot1                  & Also known as the Variable Flip-Angle in T1 (VFA)                                                                                                                                                                                                                                                                                                                                                                                                        \\
                                                                                                                          & \multicolumn{1}{l}{}     &                                                                                                                                                                                                                                                                                                                                                                                                                                                    \\
\begin{tabular}[c]{@{}l@{}}Driven Equilibrium \\ Single-Pulse Observation \\ of T2\end{tabular}                           & despot2                  &      Also known as the Variable Flip-Angle in T2 (VFA)                                                                                                                                                                                                                                                                                                                                                                                                                                              \\
                                                                                                                          & \multicolumn{1}{l}{}     &                                                                                                                                                                                                                                                                                                                                                                                                                                                    \\
\begin{tabular}[c]{@{}l@{}}(Multi-echo) Magnetisation \\ Prepared (2) Rapid \\ Acquisition Gradient-Echo(es)\end{tabular} & mprage                   & \begin{tabular}[c]{@{}l@{}}Reduces susceptibility effects including eddy currents \\ associated with metal.\end{tabular}                                                                                                                                                                                                                                                                                                                           \\* \bottomrule
\end{longtable}
\subsubsection{Medical Imaging Modality labels}

Table \ref{tab:medicalimaging} shows the different modalities of medical images classified by the energy used in the acquisition, together with the DICOM Modes that belong to the mentioned categories.\vspace{10mm}

% Please add the following required packages to your document preamble:
% \usepackage{booktabs}
% \usepackage{multirow}
% \usepackage[table,xcdraw]{xcolor}
% If you use beamer only pass "xcolor=table" option, i.e. \documentclass[xcolor=table]{beamer}
% \usepackage{longtable}
% Note: It may be necessary to compile the document several times to get a multi-page table to line up properly
\begin{ThreePartTable}
\begin{TableNotes}
\footnotesize
 \item [*] 	MRI modalities existing in BIDS and expanded in MIDS for MRI, the $<$modality\_label$>$ tags must be placed for these cases
\item [**] 	Retired modalities incorporated in the DICOM modality label
\end{TableNotes}
\begin{longtable}[c]{@{}llll@{}}
\caption{Classification of the equipment used to acquire the images, according to the medical image modality classification.}
\label{tab:medicalimaging}\\

\toprule
\multicolumn{1}{c}{\textbf{Medical Image Modality}} & \multicolumn{1}{c}{\textbf{\begin{tabular}[c]{@{}c@{}}Modality Label \\ of Medical Image\\ /mim-\\ \textless{}modality\_medical\\ \_image\_label\textgreater{}\end{tabular}}} & \multicolumn{1}{c}{\textbf{\begin{tabular}[c]{@{}c@{}}DICOM Modality,\\  See Section C.7.3.1.1.1\\ Modality (0008,0060)\end{tabular}}} & \multicolumn{1}{c}{\textbf{\begin{tabular}[c]{@{}c@{}}DICOM Modality\\  Label\\ \textless{}modality\_dicom\textgreater{}\end{tabular}}} \\* \midrule
\endfirsthead
\multicolumn{4}{c}%
{{\bfseries Table \thetable\ continued from previous page}} \\
\endhead
\bottomrule
\endfoot
\endlastfoot
                                                    &                                                                                                                                                                               & Computed radiography                                                                                                                   & cr                                                                                                                                      \\
                                                    &                                                                                                                                                                               & Bone densitometry (X-ray)                                                                                                              & bmd                                                                                                                                     \\
                                                    &                                                                                                                                                                               & X-ray angiography                                                                                                                      & xa                                                                                                                                      \\
                                                    &                                                                                                                                                                               & Digital radiography                                                                                                                    & dx                                                                                                                                      \\
                                                    &                                                                                                                                                                               & Computed tomography                                                                                                                    & ct                                                                                                                                      \\
                                                    &                                                                                                                                                                               & Intra-oral radiography                                                                                                                 & io                                                                                                                                      \\
                                                    &                                                                                                                                                                               & Mammography                                                                                                                            & mg                                                                                                                                      \\
                                                    &                                                                                                                                                                               & Videofluorography**                                                                                                                    & vf                                                                                                                                      \\
                                                    &                                                                                                                                                                               & Radio fluoroscopy                                                                                                                      & rf                                                                                                                                      \\
                                                    &                                                                                                                                                                               & Cinefluorography**                                                                                                                     & cf                                                                                                                                      \\
                                                    &                                                                                                                                                                               & Radiotherapy image (rx)                                                                                                                & rtimage                                                                                                                                 \\
                                                    &                                                                                                                                                                               & Radiotherapy plan                                                                                                                      & rtplan                                                                                                                                  \\
                                                    &                                                                                                                                                                               & RT Treatment record                                                                                                                    & rtrecord                                                                                                                                \\
                                                    &                                                                                                                                                                               & Radiotherapy dose                                                                                                                      & rtdose                                                                                                                                  \\
                                                    &                                                                                                                                                                               & Digital Fluoroscopy**                                                                                                                  & df                                                                                                                                      \\
                                                    &                                                                                                                                                                               & Panoramic X-ray                                                                                                                        & px                                                                                                                                      \\
                                                    &                                                                                                                                                                               & \begin{tabular}[c]{@{}l@{}}Radiographic imaging \\ (conventional film/screen)\end{tabular}                                             & rg                                                                                                                                      \\
\multirow{-18}{*}{General\_Radiology}               & \multirow{-18}{*}{rx}                                                                                                                                                         & Digital Subtraction Angiography**                                                                                                      & ds                                                                                                                                      \\
                                                    &                                                                                                                                                                               &                                                                                                                                        &                                                                                                                                         \\
                                                    &                                                                                                                                                                               & Magnetic resonance angiography**                                                                                                       & ma                                                                                                                                      \\
                                                    &                                                                                                                                                                               & Magnetic resonance spectroscopy-**                                                                                                     & ms                                                                                                                                      \\
                                                    &                                                                                                                                                                               & Radiotherapy image (mr)                                                                                                                & rtimage                                                                                                                                 \\
\multirow{-4}{*}{Magnetic Resonance}                & \multirow{-4}{*}{mr}                                                                                                                                                          &                                                                                                                                        &                                                                                                                                         \\
                                                    &                                                                                                                                                                               & Echocardiography **                                                                                                                    & ec                                                                                                                                      \\
                                                    &                                                                                                                                                                               & Color flow Doppler**                                                                                                                   & cd                                                                                                                                      \\
                                                    &                                                                                                                                                                               & Cystoscopy**                                                                                                                           & cs                                                                                                                                      \\
                                                    &                                                                                                                                                                               & Duplex Doppler**                                                                                                                       & dd                                                                                                                                      \\
                                                    &                                                                                                                                                                               & Intravascular ultrasound                                                                                                               & ivus                                                                                                                                    \\
                                                    &                                                                                                                                                                               & \begin{tabular}[c]{@{}l@{}}Ophthalmic axial measurements \\ (ultrasound)\end{tabular}                                                  & oam                                                                                                                                     \\
                                                    &                                                                                                                                                                               & Radiotherapy image (ultrasound)                                                                                                        & rtimage                                                                                                                                 \\
\multirow{-8}{*}{Ultrasound}                        & \multirow{-8}{*}{us}                                                                                                                                                          & Bone Densitometry (ultrasound)                                                                                                         & bdus                                                                                                                                    \\
                                                    &                                                                                                                                                                               &                                                                                                                                        &                                                                                                                                         \\
                                                    &                                                                                                                                                                               & \begin{tabular}[c]{@{}l@{}}Positron emission tomography \\ (PET)\end{tabular}                                                          & pt *                                                                                                                                    \\
\multirow{-2}{*}{Nuclear Medicine}                  & \multirow{-2}{*}{nm}                                                                                                                                                          & \begin{tabular}[c]{@{}l@{}}Single-photon emission \\ computed tomography \\ (SPECT)  **\end{tabular}                                   & st                                                                                                                                      \\
                                                    &                                                                                                                                                                               &                                                                                                                                        &                                                                                                                                         \\
                                                    &                                                                                                                                                                               & \begin{tabular}[c]{@{}l@{}}Intravascular optical coherence\\ Tomography\end{tabular}                                                   & ivoct                                                                                                                                   \\
                                                    &                                                                                                                                                                               & Autorefraction                                                                                                                         & ar                                                                                                                                      \\
                                                    &                                                                                                                                                                               & \begin{tabular}[c]{@{}l@{}}Optical coherence tomography\\ (non-ophthalmic)\end{tabular}                                                & oct                                                                                                                                     \\
                                                    &                                                                                                                                                                               & \begin{tabular}[c]{@{}l@{}}Ophthalmic axial measurements\\ (optical)\end{tabular}                                                      & oam                                                                                                                                     \\
                                                    &                                                                                                                                                                               & Ophthalmic photography                                                                                                                 & op                                                                                                                                      \\
                                                    &                                                                                                                                                                               & Ophthalmic mapping                                                                                                                     & opm                                                                                                                                     \\
                                                    &                                                                                                                                                                               & Ophthalmic tomography                                                                                                                  & opt                                                                                                                                     \\
                                                    &                                                                                                                                                                               & \begin{tabular}[c]{@{}l@{}}Ophthalmic tomography B-scan \\ volume analysis\end{tabular}                                                & optbsv                                                                                                                                  \\
                                                    &                                                                                                                                                                               & \begin{tabular}[c]{@{}l@{}}Ophthalmic tomography \\ en face\end{tabular}                                                               & optenf                                                                                                                                  \\
                                                    &                                                                                                                                                                               & Endoscopy                                                                                                                              & es                                                                                                                                      \\
                                                    &                                                                                                                                                                               & Slide microscopy                                                                                                                       & sm                                                                                                                                      \\
                                                    &                                                                                                                                                                               & General microscopy                                                                                                                     & gm                                                                                                                                      \\
                                                    &                                                                                                                                                                               & Diaphanography                                                                                                                         & dg                                                                                                                                      \\
\multirow{-14}{*}{Light}                            & \multirow{-14}{*}{light}                                                                                                                                                      & External-camera photography                                                                                                            & xc                                                                                                                                      \\
                                                    &                                                                                                                                                                               &                                                                                                                                        &                                                                                                                                         \\
                                                    &                                                                                                                                                                               & Electrocardiography                                                                                                                    & {ecg  *}                                                                                                           \\
\multirow{-2}{*}{Electrical Activities}             & \multirow{-2}{*}{elect}                                                                                                                                                       & Cardiac electrophysiology                                                                                                              & eps                                                                                                                                     \\* \bottomrule
\insertTableNotes
\end{longtable}
\end{ThreePartTable}

\subsection{Tabular files of meta-information}

\begin{figure}[]
	\centering
	\includegraphics[width=1\linewidth]{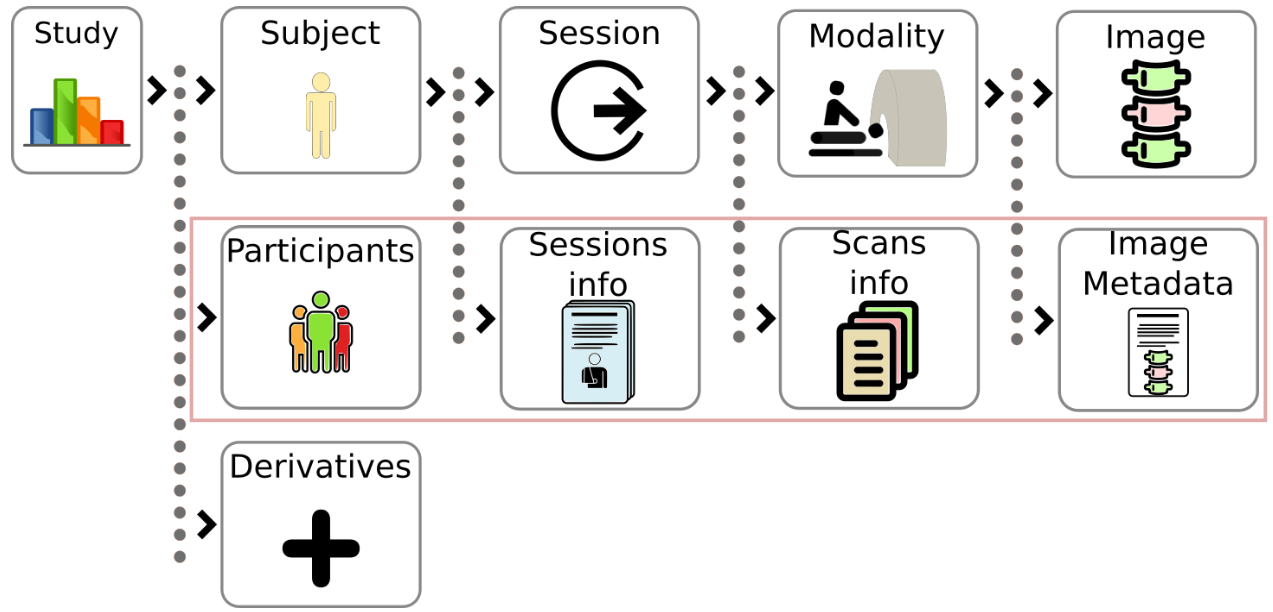}
	\caption{Diagram of folders in which each directory level has its information file.}
	\label{fig:fig3}
\end{figure}

BIDS has different levels of information organised in tabular files. These are highlighted in Figure \ref{fig:fig3} and explained below together with the improvements incorporated in MIDS.

\subsubsection{Participant description table}
This file describes the participants’ properties, such as age, handedness or sex, among others. In single-session studies, this file has one compulsory column, participant\_id, which consists of sub-$<$participant\_label$>$, followed by a list of optional columns describing the participants with only one row for each participant. This optional file is implemented in BIDS. Table \ref{tab:tab6} shows some columns of the participants.tsv description table. 

% Please add the following required packages to your document preamble:
% \usepackage{booktabs}
% \usepackage{graphicx}
\begin{table}[]
\centering
\caption{Some columns of the description table of participants include in "participants.tsv"}
\label{tab:tab6}
\resizebox{\textwidth}{!}{%
\begin{tabular}{@{}ll@{}}
\toprule
participant\_id         & \begin{tabular}[c]{@{}l@{}}REQUIRED. Tag for a patient in MIDS. Pseudonymisation may be a long and\\ confusing identification, which makes it necessary to generate a more compact\\ identifier (Participant).\end{tabular}                                                                                                                                                                                                                                    \\
                        &                                                                                                                                                                                                                                                                                                                                                                                                                                          \\
modality\_dicom             & \begin{tabular}[c]{@{}l@{}}OPTIONAL. List all of types of equipment that acquired the original data used\\to create images for this patient, DICOM (0008,0060). \end{tabular}                                                                                                                                                                                                                                                                                                                                \\
                       &                                                                                                                                                                                                                                                                                                                                                                                                                                                               \\
body\_parts             & \begin{tabular}[c]{@{}l@{}}OPTIONAL.  List all of terms for Body Part Examined in this patient, \\ DICOM (0018,0015).\end{tabular}                                                                                                                                                                                                                                                                                                                                \\
                        &                                                                                                                                                                                                                                                                                                                                                                                                                                                               \\
age                     & OPTIONAL. Age of the last session.                                                                                                                                                                                                                                                                                                                                                                                                                             \\
                        &                                                                                                                                                                                                                                                                                                                                                                                                                                                               \\
patient\_sex            & OPTIONAL. Sex of the patient.                                                                                                                                                                                                                                                                                                                                                                                                                                  \\
\multicolumn{1}{c}{...} & \multicolumn{1}{c}{...}                                             \\ \bottomrule                                                                                                                                                                                                                                                                                                                                                                                         
\end{tabular}%
}
\end{table}
\subsubsection{Session description table}
This file is optional as it provides the information referring to all the patient’s sessions, such as age at the session, procedures, or diagnoses made. The name of this file, if it is required, is ``sub-$<$\textit{participant\_label}$>$\_sessions.tsv''. This file was not described in BIDS, indeed, it is a new contribution of MIDS. Some columns of the session’s description table can be seen in Table \ref{tab:tab7}.

% Please add the following required packages to your document preamble:
% \usepackage{booktabs}
% \usepackage{graphicx}
\begin{table}[]
\centering
\caption{Some columns of the description table of sessions include in "sub-$<$\textit{participant\_id}$>$\_ sessions.tsv"}
\label{tab:tab7}
\resizebox{\textwidth}{!}{%
\begin{tabular}{@{}ll@{}}
\toprule
session\_id             & REQUIRED.                                                                                                                                                                                                                                                                                                                                                                                                                                           \\
                        &                                                                                                                                                                                                                                                                                                                                                                                                                                                     \\
radiological\_report    & \begin{tabular}[c]{@{}l@{}}OPTIONAL. Written communication between the radiologist who interprets the study\\ images and the doctor who ordered the test.\end{tabular}                                                                                                                                                                                                                                                                              \\
                        &                                                                                                                                                                                                                                                                                                                                                                                                                                                     \\
adquisition\_date       & OPTIONAL. Date of session.                                                                                                                                                                                                                                                                                                                                                                                                                           \\
                        &                                                                                                                                                                                                                                                                                                                                                                                                                                                     \\
Age                     & OPTIONAL. Age at session.                                                                                                                                                                                                                                                                                                                                                                                                                   \\
                        &                                                                                                                                                                                                                                                                                                                                                                                                                                                     \\
idc\_version            & \begin{tabular}[c]{@{}l@{}}REQUIRED BY CONDITION if “all diagnostics” or “all procedures” are filled in. \\ ICD is the acronym of the “International Classification of Diseases”.  We currently \\ use the Spanish electronic version CIE10ES \cite{martin2002cie}.\end{tabular} \\
                        &                                                                                                                                                                                                                                                                                                                                                                                                                                                     \\
all\_diagnostics        & \begin{tabular}[c]{@{}l@{}}OPTIONAL. Tag or list of tags that represents a patient diagnosis code in the ICD\\ version. \\ (e.g., {[}{[}M80.08XA{]}, {[}C34.90{]},{[}C79.51{]},{[}F17.210{]}{]} represents osteoporosis related\\ to age with current pathological fracture, vertebrae (s), initial contact for fracture\\ and other secondary diagnosis).\end{tabular}                                                                              \\
                        &                                                                                                                                                                                                                                                                                                                                                                                                                                                     \\
all\_procedures         & \begin{tabular}[c]{@{}l@{}}OPTIONAL. Tags or list of tags of the procedures in ICD version represented during\\  recognition.\\ (e.g.{[}{[}BR39ZZZ{]},{[}0BJ08ZZ{]},{[}BW25Y0Z{]},{[}BR39ZZZ{]}{]} represents magnetic \\ resonance image without contrast of the lumbar spine and other secondary probes).\end{tabular}                                                                                                                             \\
                        &                                                                                                                                                                                                                                                                                                                                                                                                                                                     \\
\multicolumn{1}{c}{...} & \multicolumn{1}{c}{...}                                                                                                                                                                                                                                                                                                                                                                                                                             \\ \bottomrule
\end{tabular}%
}
\end{table}
\subsubsection{Scan description table}
This optional file is implemented in BIDS. All data relative to the scan should be entered in this table. The table can be completed with more relevant DICOM tags for a study. The only required column as an identifier is the relative path to the image.

\subsection{Derived dataset}
Derivatives of the raw data must be kept separate from the raw data, may be under a derivatives/ subfolder in the root of the MIDS-Raw dataset folder.  To represent these derived data, the Extension Proposal BEP003 \cite{bidsDerivatives2019web} is used. 

Each pipeline has a dedicated directory under which it stores all of its outputs. There is no restriction on the directory name; however, it is recommended to use the format  $<$\textit{pipeline}$>$-$<$\textit{variant}$>$  in cases where it is anticipated that the same pipeline will output more than one variant (e.g., preprocessing-registration, segmentation-automatic, segmentation-manual, etc.).
MIDS-Derivatives filenames must follow all MIDS-Raw file naming conventions.  This inheritance principle (i.e., MIDS-Derivative inherits all BIDS-Raw rules) applies to all aspects of the spec
Every derivatives directory must include a dataset\_description.json file at the root level. $<$\textit{dataset}$>$/derivatives/$<$\textit{pipeline\_name}$>$/dataset\_description.json

This template includes the generic keys that describe the derivative data. It is the researchers’ decision to determine whether or not particular filename keys are used. 
The following is a breakdown of the generic template:

\begin{itemize}
    \item desc-<\textit{label}>: The desc keyword is a general purpose field with freeform values. To distinguish between multiple different versions of processing for the same input data the desc keyword should be used  (e.g., \_desc-manual, \_desc-UNET, etc.).
    \item label-<\textit{label}>: the label key can be used to specify a class label  when the file  belongs to a particular category or represent a single tissue class (e.g., \_label-muscle, \_label-sacrum, etc.).
    \item Suffix: The suffix indicates further details of the file's contents (e.g., \_seg, \_roi, etc.).
    \item ext: When selecting a new file format for data, open and widely used should be preferred (e.g., .png, .nifti, etc.).
\end{itemize}

\section{Software \& data}

All software for building and running the BIMCV also reading metadata of its datasets is open source and available at Github \cite{midsGithub}. Besides, the custom scripts used to combine metadata into a MIDS files structure are available in this Github.

\begin{figure}[h]
	\centering
	\includegraphics[width=0.9\linewidth]{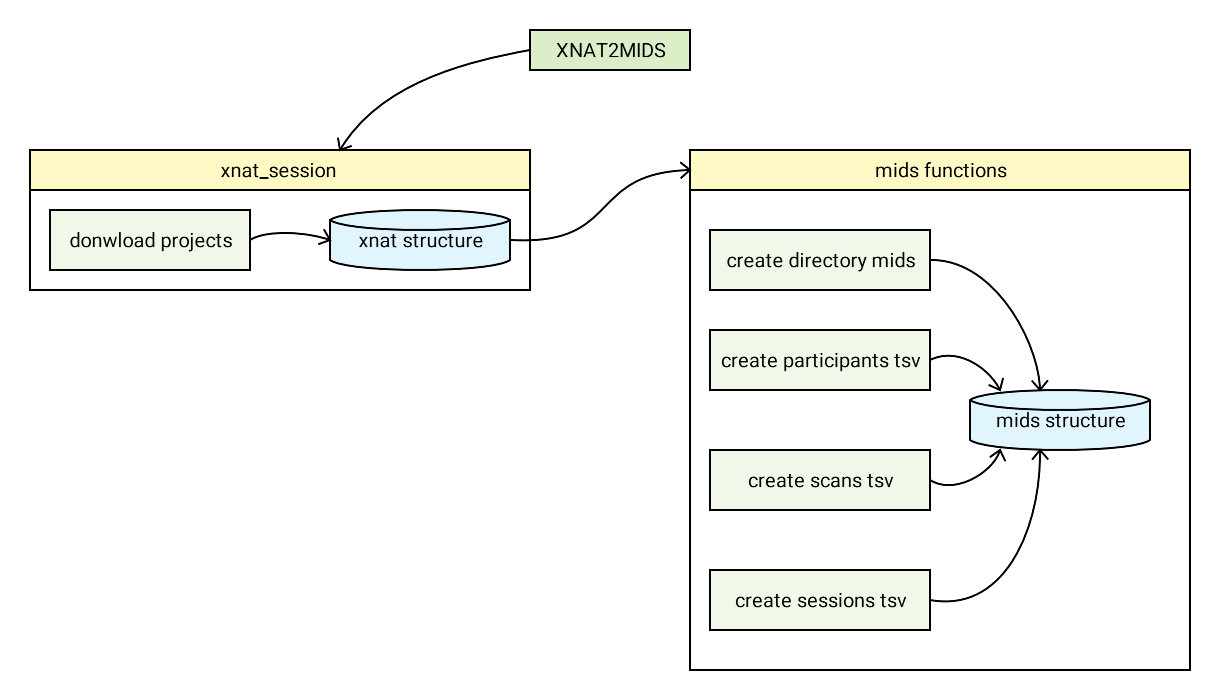}
	\caption{XNAT2MIDS execution flow.}
	\label{fig:fig4}
\end{figure}

XNAT2MIDS is a software written in Python 3 and can be found at the public Github repository listed above. This software allows the users of an XNAT platform to connect to their assigned projects and download the projects of interest in MIDS format. Figure \ref{fig:fig4} shows the execution flow in which a XNAT session is generated. It can store all the requested projects and, once saved in a temporary directory, the program generates the general MIDS structure and its tabular metadata.

\begin{figure}[]
	\centering
	\includegraphics[width=1\linewidth]{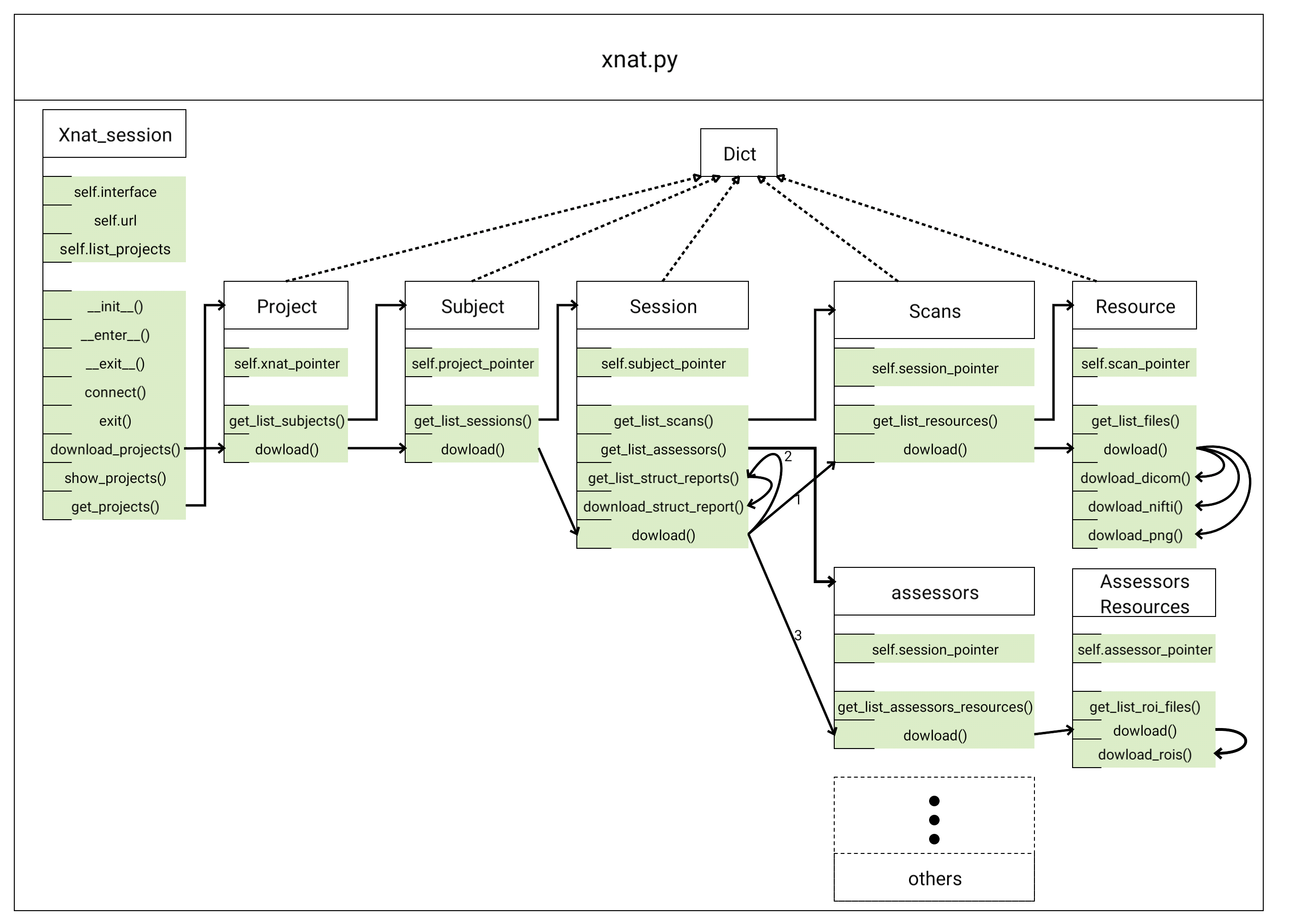}
	\caption{Class diagram for downloading projects in XNAT.}
	\label{fig:fig5}
\end{figure}

It was decided to make the download based on requests with the Rest API of the web platform for the connection with XNAT. This application ensures that the projects downloaded are saved to the disk without transmission errors. Large projects, contrary to standard web methods, will be interrupted during the download. The download software is made up of classes that are responsible for transmitting information to the lower level until each image is downloaded. The structure of  XNAT code can be seen in Figure \ref{fig:fig5}.

In a future project, we are going to use Machine Learning techniques to improve the auto-tagging of the MIDS structure, for example, by tagging the MRI type or the acquired anatomical part using exclusively DICOM features such as the acquisition parameters or image features through computer vision methods.

We also intend to include other software, such as the MIDS validator, enabling it to accept MIDS structure as an extension. All the datasets described in this proposal are available in Github.

 This directory structure was applied to the data in the medical image bank of the Valencian community in the BIMCV COVID19 + project \cite{vaya2020bimcv}. This project is a large dataset of the Medical Imaging Databank in BIMCV with chest X-ray images CXR (CR, DX) and computed tomography (CT) imaging of SARS-CoV-2 (COVID 19). The dataset can be downloaded from http://bimcv.cipf.es/bimcv-projects/bimcv-COVID-19 in the MIDS structure.

\section{Discussion}

MIDS is proposed as an extension of BIDS to include datasets with medical images of different body parts obtained by different methods. It adds a new element, the Session description table, to the BIDS structure, in which relevant information on the session is stored. 

Due to the high diversity of use cases that fit in the MIDS standard, MIDS accepts the BIDS key-value indexes if they are needed in image types or body parts that differ from those accepted by BIDS. This will be expanded in the future by including use cases supervised by experts to define these indexes.

The organisation by type of physical principle for each image makes the structure better categorised and therefore more intuitive for the user. Another advantage of separating by energy is the possibility of categorising all the possible methods of obtaining an image, which gives it the flexibility to catalogue images from new devices.

The specialisation of the type of image enhanced from its sequence is essential when training artificial intelligence models. For example, a T2w image in Spin Echo sequence is not the same as in Gradient Echo sequence since it causes the intensity to change for the same tissues at the pixel intensity level, so it would be wrong to classify them as the same type of image.

In the current MIDS version, only the ICD codification is used to describe the diagnosis and procedures performed in/at the session. In future versions we hope to expand this to other codifications such as the Unified Medical Language System (UMLS), specifically SNOMED-CT, so that researchers could use the standard that best fits their needs or allow the definition of different codifications in the same file to enrich the information displayed.

This type of standard provides a common structure for many projects, which can be helpful in artificial intelligence projects where these images can be used to train models that absorb the necessary knowledge to solve any problems that may arise. When it comes to obtain results, it can also be applied to automatically know where to place them in this structure.

BIDS/MIDS is part of a collaboration with the DeepHealth EU project in the guidance and construction of a fully anonymised population medical imaging data lake structure (according to the guidelines “Opinion 05/2014 on Anonymisation Techniques” and CEN/ISO standards). This data lake structure will allow the community to massively store imaging data in a native format to which MIDS will contribute with an improvement in the data curation process. This will provide a clear and simple structure both for the users and the software using these data.

%\bibliography{references}  %%% Remove comment to use the external .bib file (using bibtex).
%%% and comment out the ``thebibliography'' section.

%%% Comment out this section when you \bibliography{references} is enabled.
\bibliographystyle{unsrt}
%\nocite{*}
%\bibliographystyle{unsrt}
\inputencoding{latin2}
\bibliography{mids}
\inputencoding{utf8}
\newpage

%\section{Appendix 1: Figures of MIDS}

\paragraph{\Large \textbf{Acknowledgement}}
This work is first and foremost an open and free contribution from the authors
in the working group with support from the Regional Ministry of Innovation,
Universities, Science and Digital Society grant awarded through decree 51/2020
by the Valencian Innovation Agency (Spain) and Regional Ministry of Health in
Valencia Region. This research is also supported by the University of Alicante’s
UACOVID-19-18 project.

This project was funded by a grant from the Generalitat Valenciana (Covid\_19-SCI). Part of the infrastructure used has been cofunded by the European Union through the Operational Program of the European Fund of Regional Development (FEDER) of the Valencian Community 2014-2020. The Medical Image Bank of the Valencian Community (BIMCV) was partially funded by the European Union’s Horizon 2020 Framework Programme under grant agreement 688945 (Euro-BioImaging PrepPhase II).

This article describes work undertaken in the context of the DeepHealth project, “Deep-Learning and HPC to Boost Biomedical Applications for Health” (\href{https://deephealth-project.eu/}{https://deephealth-project.eu/}) which has received funding from the European Union’s Horizon 2020 research and innovation programme under grant agreement No 825111”. The contents of this publication reflect only the author’s view, can in no way be taken to reflect the views of the European Union and the Community is not liable for any use that may be made of the information contained therein.
ETHICS APPROVAL

Finally, we also thank the entire BIDS team for providing us with the necessary tools to give visibility to this proposal and Robert Oostenveld, Gustav Nilsonne, Stefan Appelhoff and Thomas Nichols for their contributions to improve the proposal.
\paragraph{\Large \textbf{Ethics approval}}
The study was approved by the local institutional ethics committee DGSP-CSISP NÚM. 20190503/12. AVAILABILITY OF DATA.
\paragraph{\Large \textbf{Rights and permissions}}
Which permits use, sharing, adaptation, distribution and reproduction in any medium or format, as long as you give appropriate credit to the original author(s) and the source, provide a link to the Creative Commons license, and indicate if changes were made. The images or other third party material in this article are included in the article's Creative Commons license, unless indicated otherwise in a credit line to the material. If material is not included in the article's Creative Commons license and your intended use is not permitted by statutory regulation or exceeds the permitted use, you will need to obtain permission directly from the copyright holder.
\end{document}